\documentclass[12pt]{article}

\input epsf.sty
\input colordvi.sty
\usepackage[]{graphicx}
\newcommand{\bright}{\begin{flushright}}
\newcommand{\eright}{\end{flushright}}
\newcommand{\beq}{\begin{equation}}
\newcommand{\eeq}{\end{equation}}
\newcommand{\beqa}{\begin{eqnarray}}
\newcommand{\eeqa}{\end{eqnarray}}
\newcommand{\half}{\frac{1}{2}}

\newcommand{\Lsim}{\mbox{\raisebox{-.3em}{$\stackrel{<}{\sim}$}}}
\newcommand{\Lmd}{\Lambda}
\newcommand{\lmd}{\lambda}
\newcommand{\MP}{M_{\rm P}}
\newcommand{\nnb}{\nonumber}
\newcommand{\reflef}{(\ref}

\newcommand{\bbox}{\sqcap\hspace{-.9em}\sqcup}
 
\begin{document}
%\vspace{-1em}
\begin{center}
%\mbox{}\\[-3.5em]
%\bright
%/gnga/nest4bc4.tex (6 sep 15)\today
%\eright
%\mbox{}\\[-1.3em]
{\Large\bf A new estimate of the mass of the gravitational scalar field for Dark Energy }
\mbox{}\\[0.6em]
{\large\bf Yasunori Fujii}\\
Advanced Research Institute for Science and Engineering, Waseda University, Okubo, Shinjuku-ku, Tokyo, 169-8555, Japan\\[-.2em]
%\mbox{}\\[.01em]
Email fujiitana@gmail.com\\[.15em]
%mbox{}\\[.1em]
%{\small  A contribution to the book {\it At the Frontier of  Spaceitime -- Scalar-Tensor Theory, \\Bells inequality, Mach's Principle, Exotic Smoothness } dedicated to the 80th \\Birthday of Prof. Carl Brans, Springer, 2016}

%\mbox{.5em}\\

\begin{abstract}

A new estimate of the mass of the pseudo dilaton  is offered by following the fundamental nature 
%faithfully 
that a massless Nambu-Goldstone boson, called a dilaton, in the Einstein frame acquires a nonzero mass through the loop effects which occur with the Higgs field in the relativistic quantum field theory as described by poles of $D$, spacetime dimensionality off the physical value $D=4$.  Naturally the technique of dimensional regulairzation is fully used to show this pole structure to be suppressed to be finite by what is called a Classical-Quantum-Interplay, to improve our previous attempt.  Basically the same analysis is extended to derive  
%non-Newtonian potential 
also the coupling of a pseudo dilaton to two photons.
\end{abstract}
\end{center}

\section{Introduction}

We have developed our own version of the Scalar-Tensor theory (STT) \cite{yfkm,Entr}  due originally to  Jordan \cite{jordan}, also to Brans and Dicke \cite{BD},  but now with the unique feature that we are then allowed to be free from the  possible fine-tuning problem in understadning the small size of a cosmological constant (CC), or the dark energy (DE), fitted to the observed accelerating universe \cite{obs}.  Today's value of CC $\sim t_0^{-2}$ with $t_0$ the present age of the universe, is this small simply because we are this old cosmologically.\footnote{We use the reduced Planckian unit system defined by $c=\hbar=\MP(=(8\pi G)^{-1/2})=1$ .  The units of length, time and energy in conventional units are given by $8.10 \times 10^{-33}{\rm cm}, 2.70\times 10^{-43}{\rm sec}, 2.44 \times 10^{18}{\rm GeV}$, respectively.  As an example of the converse of the last entry, we find $1 {\rm GeV}=2.44^{-1}\times 10^{-18}$ in units of the Planck energy.  In the same way, the  present age of the universe $t_0 \approx 1.37\times 10^{10}{\rm y}$ is $10^{60.2}$ in units of the Planck time.   }

This approach is an outgrowth, in retrospect, of a conceptual attempt based on a simple $\Lmd$ cosmology for the radiation-dominated universe \cite{dolg}. According to an attractor and asymptotic solution, the Jordan frame (JF), with a variable $G$, describes unrealistically a static universe, while  the Einstein frame (EF, with subscript $*$), with $G_*$ kept constant, provides fortunately with an expanding universe, $a(t_*)\sim t_*^{1/2}$,  hence accepted as the physical frame.  Also the scalar field density interpreted as dark energy density falls off like $t_*^{-2}$, from which follows the scenario of a decaying cosmological constant, as emphasized above.  On the other hand, however, starting off by assuming a conventional mass term in JF, we come to finding that the microscopic fundamental particles, including an electron, with their masses which {\em fall off} like $\sim a_*(t_*)^{-1}\sim t_*^{-1/2}$ in EF.  This is totally in conflict with today's view on the astronomical measurements; cosmological size is measured in reference to the microscopic units of length provided by the inverse mass of the microscopic particles.  Also we have no way of detecting any variation of units themselves, implying their constancy in the physical frame, as we once called the {\em Own-unit-insensitivity principle}(OUIP) \cite{yf118}, which ultimately derives the receding speeds of distant objects in terms of the red-shifts of the observed  atomic spectra.

From these arguments we view the theory of STT to face a serious flaw when it meets the microscopic physics.  Sometime ago we came across \cite{yfkm,Entr} that this flaw can be avoided miraculously in terms of global scale invariance broken spontaneously with the scalar field playing the role of a massless Nambu-Goldstone (NG) boson \cite{nambu,gold,NJL},
 called {\em dilaton}, which, like many other examples of NG bosons, would acquire a nonzero mass hence a {\em pseudo dilaton}.  We further suggested a tentative estimate of its mass-squared;  $\mu^2 \sim m_{\rm q}^2 M_{\rm ss}^2/\MP^2 \sim (10^{-9}{\rm eV})^2$, with $m_{\rm q}$ for the averaged quark mass, while $M_{\rm ss} \sim 10^3 {\rm GeV}$ for the supersymmetry mass scale to be prepared for  
the quadratic cutoff of the self-energy of a scalar field.\footnote{ This mass corresponds approrximately to the force-range $\sim 100 {\rm m}$, 
 related to the suggested non-Newtonian force, as was discussed in \cite{yfnat}, for example.}

Now in the current article, we are going to replace quarks by the Higgs field as an  origin for the masses of fundamental particles, also with an improved technique for the  mass acquisition mechanism applied uniquely to the dilaton.  The new numerical estimate based on the Standard Model  (SM) turns out to result in $\mu$ somewhat heavier than our previous estimate, still basically more or less in the same range of the  order of magnitude, remaining responsible for the experimental searches for the DE \cite{DEexp} of the accelarating universe through $\gamma\gamma$ scattering.

On  the theoretical side, as we also point out, the two different concepts, scaling behavior and infinities in the qantized field theory, are described by a single common variable,  the spacetime dimensionality assumed to be continuous off the physical value 4.  
On the basis of dimensional regularization (DR) technique \cite{gth}, the formulation is then not only simple and straightforward  but also  continued smoothly from  the {\em first half} of the spontaneouly broken scale invariance.

The same mechanism applies also to the coupling of a pseudo dilaton to two photons.  These experiences prompt us to present our phenomenological results even, for the moment, only with a preliminary study of the effects of the conventional renormalization procedure.

%%%%%%%%%%%%%%%%%%%%%%%%%%%%%

\section{Basic equations}

We start with the basic Lagrangian
%%%%%%%%%
\beq
{\cal L}=\left\{
\begin{array}{l}
\sqrt{-g}\left( \half\xi\phi^2 R-\half\epsilon g^{\mu\nu}\partial_\mu\phi\partial_\mu\phi +L_{\rm matter} -\Lmd    \right), \\[.6em]
\sqrt{-g_*}\left( \half R_*-\half g_*^{\mu\nu}\partial_\mu\sigma\partial_\mu\sigma +L_{\rm *matter} -\Lmd \exp (-4\hat{\zeta}\sigma)    \right),
\label{nest_1}
\end{array}
\right.
\eeq
for $\Lmd >0$ and $\xi >0$ by avoiding antigravity,  expressed both in JF and EF, respectively, where the factor $\Omega$ for the conformal transfromation satisfies
\footnote{Symbols $\varphi, \omega$ used in the original references \cite{jordan,BD} are now re-expressd by our more convenient ones; $\varphi =(1/2)\xi \phi^2, 4\omega =\epsilon\xi^{-1}$.}
\beq
g_{\mu\nu}=\Omega(x)^{-2}g_{*\mu\nu},\quad \phi=\hat{\xi}^{-1/2}\Omega,\quad\mbox{with}\quad \Omega =\exp(\hat{\zeta}\sigma),
%\zeta^{-2}\equiv 6+\epsilon \xi^{-1} >0,
\label{nest_2}
\eeq
where
%%%%%
\beq
\hat{\xi}=\xi\MP^{-2},\quad \hat{\zeta}=\zeta\MP^{-1},\quad\mbox{with}\quad \zeta^{-2}\equiv 6+\epsilon \xi^{-1} >0.
\label{nest_2_2}
\eeq
The symbol $\phi$ implies the gravitational scalar field in JF with  $\epsilon =\pm 1$, or 0.\footnote{$\epsilon =0$ implies $\zeta^2=1/6$ corresponding to the coefficient $a=1/3$ of the scalar component of the combined potential  \cite{yfnat,Ohanl}.  We choose $\eta_{00}=\eta^{00}=-1$. }  Notice that $\phi$, even with an apparently ghost nature with $\epsilon =-1$, has been brought to be mixed to the spinless portion of $g_{\mu\nu}$ through the nonminimal coupling term, the first term on RHS of the upper line of \reflef{nest_1}), to emerge as a canonical nonghost scalar field $\sigma$ in EF under the condition specified at the last of \reflef{nest_2_2}).

From \reflef{nest_1}) we derive the cosmological equations in each of radiation-dominated JF and EF, also in the standard spatially flat Robertson-Walker metric with the matter density approximated by a uniform distribution, finding the attractor and asymptotic solutions.\footnote{For details see \cite{yfkm,Entr,MaedaFujii}.}   This is the way we reach the static universe in JF and the expanding universe in EF, as stated in Sec. 1.

During the calculation, we derived the asymptotic relation for the matter density $\rho $ in  JF;
%%%%%%
\beq
\rho =-3\Lmd \frac{2\xi+\epsilon}{6\xi+\epsilon}.
\label{nest_2a}
\eeq
Using $\Lmd >0$ and $\xi >0$ as stated immediately following \reflef{nest_1}) before also the last condition of \reflef{nest_2_2}), 
we find that the obvious condition $\rho >0$ results only if 
%%%%%%%%
\beq
\epsilon = -1, \quad\mbox{hence} \quad \frac{1}{6}<\xi<\frac{1}{2},\quad\mbox{and}\quad \frac{1}{4} <\zeta^{2}< \infty, 
\label{nest_2b}
\eeq
as will be used later.  We also come to find
%%%%%%%
\beq
\phi \rightarrow t, \quad\mbox{or}\quad \Omega\rightarrow t, \quad\mbox{as}\quad t \rightarrow \infty.
\label{nest_2c}
\eeq

Many of the important features in these approximate solutions are taken over to the more exact numerical solutions, which include such a unique complication like the occasional step-like falling-off behavior of the scalar-field density, as was discussed with the help of a supporting assumption, in SubSecs 5.4.1-2 of \cite{yfkm} and SubSecs 6.1-2 of \cite{Entr}, thus allowing the wording, a {\em decay}ing cosmological {\em constant}, acceptable semantically. But more urgently, we re-emphasize briefly how OUIP is observed  by the spontaneously broken scale invariance.

The crucial point is that we are supposed to start with the {\em interaction} term in JF;
%%%%%%
\beq
-{\cal L}_{\rm I} =\half \sqrt{-g}\;h\phi^2\Phi^2,
\label{mmas1_0060}
\eeq
instead of the conventional mass term $-{\cal L}_m =\sqrt{-g}(1/2)m^2\Phi^2$, applied to an example of the real scalar-field matter $\Phi$, where $h$ is a dimension{\em less} coupling constant, hence indicating {\em scale invariance} in JF.  The conformal transformation to EF yields
%%%%%%%%%
\beq
-{\cal L}_{\rm I} =\half \sqrt{-g_*}\Omega^{-4}h\hat{\xi}^{-1}\Omega^2\Omega^2 \Phi_*^2 =\half \sqrt{-g_*}m_*^2\Phi_*^2,\quad\mbox{with}\quad \Phi=\Omega \Phi_*,\quad  m_*^2=h\hat{\xi}^{-1}.
\label{mmas1_0061}
\eeq
Notice that all of the $\Omega$'s cancel each other, hence leaving a truly {\em constant} $m_*$.  The last result might be combined with $\xi \sim {\cal O}(1)$ as obtained from the second of \reflef{nest_2b}) to find $h\sim {\cal O}(m_*^2/\MP^2)$, which will be inherited basically to the more realistic but more complicated model for the Higgs field as will be developed soon later.

In this connection, we also notice that \reflef{mmas1_0060}) fails to observe a premise that $\phi$ be decoupled from the matter Lagrangian, as emphasized by Brans and Dicke  \cite{BD} who realized this to be the simplest way to implement Weak Equivalence Principle (WEP) as one of the most important features in the macroscopic and classical gravity.\footnote{The amplitude through the nonminimal coupling term observes the same tensor coupling as in General Relativity.}

Now facing the microscopic and cosmological gravity, we might try an attempt beyond their premise.  This is the reason why we come to  \reflef{mmas1_0060}), showing remarkably the global scale invariance {\em taking} $\phi$ {\em into account}.  In a sense, we exploit the scale invariance of the whole STT terms in JF Lagrangian except for $\Lmd$.

From none of the mass in \reflef{mmas1_0060}) we have created the mass {\em spontaneously}.  In fact the mass dimension has been smuggled through $\hat{\xi}^{-1/2}$ as a VEV of $\phi$,  following the second of \reflef{nest_2}).  The spontaneous nature might also be better interpreted by computing the Noether current of the scale transformation;\footnote{The special form of the first in  the following, coreponding to $g_{\mu\nu}\rightarrow \Omega^2 g_{\mu\nu}$, has been selected, because, unlike another type of the coordinate transformation $x^\mu \rightarrow \Omega x^\mu$ or $\delta x^\mu = \ell x^\mu$, making it straightforward to be applied to the expanding universe with the 3-space simply uniform without any particular origin.  See \cite{SS}, for example, on introducing $\delta x^\mu /x^2 = \beta^\mu$.   }
%%%%%%%%%
\beq
\delta g_{\mu\nu}=2\ell g_{\mu\nu},\quad \delta\phi =-\ell \phi, \quad  \delta\Phi =-\ell \Phi,
\label{mmas1_000}
\eeq
which derives,
\beq
J^\mu =\half \sqrt{-g}g^{\mu\nu}\partial_\nu \left( \xi\zeta^{-2}\phi^2+\Phi^2  \right),
\label{mmas1_0050}
\eeq
an exact result, as detailed in Appendix M of \cite{yfkm}. 
We then re-express RHS now in EF.  In this process, we obtain $\bbox_*\hspace{-.2em}\Phi_*^2$ on RHS, then by using the field equation of $\Phi_*$ coming to find the contributon from $\Lmd_*=\exp(-4\hat\zeta\sigma)\Lmd$.  In the rest of the present article, however, we focus upon the epochs in which $\Lmd_*$ remains rather smaller than the ordinary matter as was demonstrated by our realistic solutions in \cite{yfkm,Entr}, in the reasonable past from today.  Then we may ignore $\Lmd_*$  approximately, also impose the conservation law $\partial_\mu J_*^\mu =0$, thus obtaining 
%%%%%%%%%%
\beq
\zeta^{-1}\sqcap\hspace{-.9em}\sqcup_*\sigma =-\left( m_*^2\Phi_*^2 +g_*^{\mu\nu} \partial_\mu\Phi_* \partial_\nu\Phi_* \right),
\label{mmas1_005c}
\eeq
where LHS comes directly from the first term on RHS of \reflef{mmas1_0050}), hence vreaching the massless nature of $\sigma$, to be called a {\em dilaton}, precisely as had been shown by Nambu \cite{nambu}.  The occurrence of the  massless dilaton is expected to survive the approximations mentioned above.

We now leave the {\em first half} of the scenario of spontaneously broken scale invarinace, entering its {\em second half} in which the massless dilaton grows into
 a  {\em massive pseudo dilaton}.  For this purpose, we start with the spacetime dimensionality $D$ off, but close to, the physical value 4.  We also re-interpret the above $\Phi$ now as the Higgs field supposed to provide with the origin of the masses of all the microscopic fields.  In fact the familiar {\em Mexican-Hat} potential is shown to inherit the core of the 4-dimensional scale-invariance within the realm of STT.

We then extend \reflef{mmas1_0060}) to 
%%%%%%%
\beq
-{\cal L}_H=\sqrt{-g}\left( \half h\phi^2\Phi^2 +\frac{\lmd}{4!}\Phi^4  \right)=\sqrt{-g_*}\Omega^{D-4}\left(\half \tilde{m}^2\Phi_*^2  +\frac{\lmd}{4!}\Phi_*^4  \right),\quad\mbox{with}\quad \tilde{m}^2=h\hat{\xi}^{-1},
\label{mmas2_6}
\eeq
where $\Omega$ in the second of \reflef{nest_2}) and  
 \reflef{mmas1_0061}) are both replaced by $\Omega^{d-1}$ with $d=D/2$. Then together with $\sqrt{-g}=\Omega^{-D}\sqrt{-g_*}$, we come to find an overall multiplier on RHS of \reflef{mmas2_6});
%%%%%%%%
\beq
\Omega^{-D}\left(\Omega^{d-1})\right)^4 =\Omega^{-D+2D -4}=\Omega^{D-4}.
\label{mmas2_61}
\eeq
Obviously, adding the quartic self-coupling of $\Phi^4$ to \reflef{mmas1_0060}) leaves our experiences of the scaling behaviors and spontaneous symmetry breaking nearly unchanged.

Following the well-known procedure, we shift the origin by the VEV $v$;
%%%%%%
\beq
\Phi_*=v+\tilde\Phi. 
\label{mmas2_7}
\eeq
By requiring the absence of the term linear in $\tilde{\Phi}$, we arrive  finally at \footnote{The symbol $m$ in the second equation is the mass in EF, to be better denoted by $m_*$.  To avoid too much notational complications in the following equations, however, we continue to use $m$ without the subscript $*$ for the observed  mass for the Higgs mass in the whole subsequent part of the article.   }
%%%%%%%%
\beq
-{\cal L}_{\rm H}= \exp\left( 2\hat{\zeta} (d-2)\sigma  \right)\sqrt{-g_*}{\cal V}, \quad\mbox{with}\quad {\cal V}=
\half m^2\tilde{\Phi}^2 +\half m\sqrt{\frac{\lmd}{3}}\tilde{\Phi}^3+\frac{\lmd}{4!}\tilde{\Phi}^4 ,
\label{mmas2_15}
\eeq
where $m$ defined by
%%%%%%
\beq
m^2 =-2\tilde{m}^2=\frac{\lmd}{3}v^2, 
\label{mmas2_10d}
\eeq
is the observed mass of the Higgs field, $1.26 \times 10^2 {\rm GeV}$ \cite{hhgs}.

In this way we define the dynamics of $\tilde{\Phi}$ and $\sigma$, where the exponential factor comes from \reflef{mmas2_61}), substituted from the third of \reflef{nest_2}), while ${\cal V}$ is the Higgs potential.\footnote{This potential $\cal V$ is shown to agree with the relevant part of the SM.  See (87.3) of \cite{sred}, for example. His $V(\varphi)=(\lmd_r/4)(\varphi^\dagger \varphi -(v^2/2))^2$ is reproduced precisely by our \reflef{mmas2_15}) by choosing
 $\sigma =0, \hspace{.5em}\lmd_r/4=\lmd /4!$, and $\varphi =(1/\sqrt{2})(v+\tilde{\Phi})$ only for the single component, with another component vanishing, corresponding to his (87.13). Notice also that a special relation chosen between the two terms in the parentesis on RHS of $V(\varphi)$ above has the same effect of the term linear in $\tilde{\Phi}$ removed, which we required in the sentence just prior to the foregoing footnote 8. }    We have ignored the possible vacuum terms as well as part of CC, in accordance with what we stated before between \reflef{mmas1_0050}) and \reflef{mmas1_005c}).  As we consider, the current process of mass creation and the pseudo dilaton belongs to a {\em local physics} expected not to be affected seriously by CC, or DE.

To be noticed more explicitly, the field $\sigma$ occurs {\em only in association} with $d-2 = (D-4)/2$ separated from what is called the Higgs potential $\cal V$.  In other words, the dilaton $\sigma$ might appear to be present only for $D \neq 4$.  This by no means implies that $\sigma$ is entirely outside our realistic concern at $D=4$, because another part $\cal V$ contains infinities at $D=4$, as will be shown shortly.  We should take the same attitude as  in  DR that we keep $d\neq 2$ during the computation until we come back to the physical value $d=2$ only at the very end of the calculation.

For more details of the calculation, we apply the expansion
%%%
\beq
\exp(2\hat{\zeta}(d-2)\sigma) \approx 1-2\hat{\zeta}(2-d)\sigma.
\label{mmas2_61z}
\eeq
On the other hand, the  potential $\cal V$, representing a collection of the field $\tilde {\Phi}$,   might develop certain Feynman diagrams  which may happen to induce closed loops of $\tilde{\Phi}$ then exhibiting divergences. The simplest 1-loop divergence is described by
%%%%%%%%
\beq
\Gamma(2-d)\sim (2-d)^{-1},\quad \mbox{as }\quad d\rightarrow 2,
\label{mmas2_61a}
\eeq
according to the technique of DR.

By substituting \reflef{mmas2_61z}) and \reflef{mmas2_61a}) into the first of \reflef{mmas2_15}), we  find a product like $\zeta(d-2)\sigma \Gamma(2-d)$, for which we apply the relation
%%%%%%%%%
\beq
(2-d)\Gamma (2-d) \rightarrow 1, \quad \mbox{as }\quad d\rightarrow 2,
\label{mmas2_61b}
\eeq
to be called a {\em Classical-Quantum Interplay} (CQI), which connects a classical factor $2-d$ to $\Gamma (2-d)$ obviously representing a quantum nature.  The above result implies that $\sigma$ happens to pick up a $\tilde{\Phi}$-loop, which thus plays a major role in the same way as the cutoff conjectured by Nambu and Jona-Lasinio \cite{NJL} for the study of the pion-nuleon system, thus re-discovered in a somewhat different but closely related context.

Notice also that we no longer suffer from infinities, as far as the relevant processes are dominated by CQI. This might be, however, related to the Brans-Dicke premise as was discussed following \reflef{mmas1_0061}).  The pole structure \reflef{mmas2_61a}) should apply only to such fundamental fields like $\Phi$, or quarks and leptons, but not to composite particles like hadrons.  In this sense, the WEP violation effect due to the occurrece of $\phi$ in the matter Lagrangian tends to be smaller, but might need more detailed analyses before reaching the final comparison with the observations.

In the next Section we are going to discuss how this crucial relation, CQI, can be used naturally in calculating the mass of the pseudo dilaton.

\section{Computing the mass of the pseudo dilaton}
\setcounter{equation}{0}

In order to derive the mass  $\mu$ of $\sigma$, we first consider the simplest form of the $\sigma$ self-energy (SE) part, as shown in the upper line of Fig. 1, where the dotted and solid lines are for $\sigma$ and $\tilde{\Phi}$, respectively.  Each vertex is read out  from the second term on RHS of \reflef{mmas2_61z}) times the   first term in $\cal V$ of \reflef{mmas2_15}), deriving the effective vertex part;
%%%%%
\beq
g_0 =-2\hat{\zeta}(2-d)m^2.
\label{mmas2_70}
\eeq
The loop integral of $\tilde{\Phi}$ gives\footnote{Some of the details of the required integrals will be found in Appendix N of \cite{yfkm}. }$^{,}$\footnote{ Strictly speaking, the denominators should be $((k+q/2)^2+m^2)((k-q/2)^2+m^2)$, where $q$ is for the momentum of the size of $\sim \mu$. Since we finally find $\mu$ negligibly smaller than $m$ as in  \reflef{mmas_37}), we might justify the approximate computation as in \reflef{mmas2_71}). The same kind of approximation applies to almost any of the loop inetegrals to be encountered in the following of the present article.}
%%%%%%%%%
\beq
\int d^Dk \frac{1}{(k^2+m^2)^2}=i\pi^2\left( m^2\right)^{d-2}\frac{\Gamma (d)\Gamma (2-d)}{\Gamma (2)} \approx i\pi^2 \Gamma(2-d).
\label{mmas2_71}
\eeq
The whole contribution is
%%%%%%%%
\beq
\sim g_0^2i\pi^2 \Gamma (2-d)\sim 4m^4\zeta^2 (2-d)^2 \Gamma (2-d)\sim 2-d \rightarrow 0,\quad \mbox{as }\quad d\rightarrow 2,
\label{mmas2_72}
\eeq
since the pole $(2-d)^{-1}$ has been over-cancelled  by $(d-2)^2$ in accordance with CQI.  As a lesson, a finite nonzero result will occur only if the $\tilde{\Phi}$ in ${\cal V}$ must include 2-loop divergences.

%%%%%%%%%%%%%%%%%%%%%%%%%%%%%%% ffg %%%%%%%%%%%%%%%%

\begin{figure}[h]
\begin{center}
\includegraphics[width=150mm]{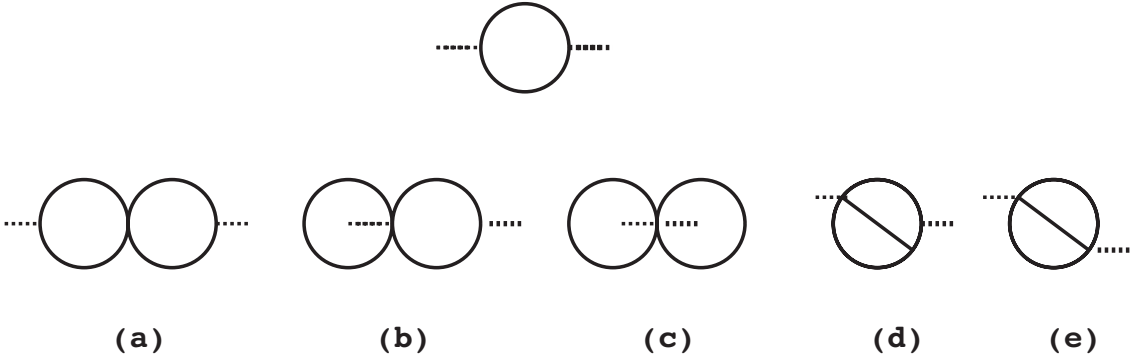}
\end{center}
\caption{In the upper line we show the simplified 1-loop SE part of $\sigma$, to start with.   Dotted and solid lines are for the $\sigma$ and the shifted Higgs field $\tilde{\Phi}$, respectively.   In the lower line, we illustrate the 2-loop amplitudes to which CQI can be applied.    They are different  from each other in the ways the dotted lines couple to different parts of the three terms in the potential $\cal V$ on RHS of \reflef{mmas2_15}).
}
\end{figure}

We now try to extend the argument to construct 2-loop amplitudes for the $\sigma$ mass term, as illustrated by (a)-(e) in the lower line of Fig. 1.  First in the diagram (a), we show a simply doubled 1-loop diagrams, in fact by connecting the simplest one to another diagram formed by the first term and the third term of ${\cal V}$ in \reflef{mmas2_15}).  By computing explicitly, we do reach two poles resulting in the nonzero result; 
%%%%%%%%%
\beqa
{\cal J}_a &=& \left[ -i(2\pi)^4  \right]^3 \left[i(2\pi)^4 \right]^{-4}\lmd m^4\left[ 2\hat{\zeta}(2-d)\sigma  \right]^2
\int d^D k \frac{1}{(k^2+m^2)^2}\int d^D k' \frac{1}{(k'^2+m^2)^2}\nnb\\
&=&i(2\pi)^{-4}(i\pi^2)^2 \lmd \hat{\zeta}^2 4m^4 \sigma^2=
 -i\frac{1}{4}\lmd\hat{\zeta}^2 m^4\sigma^2, \quad\mbox{as}\quad d\rightarrow 2.
\label{mmas2_16}
\eeqa

The combination in (b) is basically the same as in (a),   different only where
 one of the dilaton lines reaches $\cal V$, thus causing a {\em linear} denominator.  Still due to 
%%%%%%%%%
\beq
\int d^Dk' \frac{1}{k'^2+m^2}=i\pi^2 (m^2)^{d-1}\frac{\Gamma (d)\Gamma (1-d)}{\Gamma (3-d)} =-i\pi^2 m^2\Gamma(2-d),
\label{mmas2_71a}
\eeq
somewhat different from \reflef{mmas2_71}), though, we find the result, which happens to be the same as \reflef{mmas2_16});
%%%%%%%%%
\beqa
{\cal J}_b&=&\left[ -i(2\pi)^4  \right]^2  \left[i(2\pi)^4 \right]^{-3}
\lmd m^2 \left[ 2\hat{\zeta}^2 (2-d)\sigma \right]^2 
 \int d^Dk' \frac{1}{k'^2+m^2}\int d^Dk\frac{1}{(k^2+m^2)^2}\nnb\\
&=&i\frac{1}{4}\pi^{-4} \lmd m^4 (i\pi^2)^2 \hat{\zeta}^2\sigma^2=-i\frac{1}{4}\lmd \hat{\zeta}^2 m^4 \sigma^2=J_a.
\label{mmas_17}
\eeqa

The same result will follow for the diagram in (c);
%%%%%%
\beqa
{\cal J}_c&=&-i(2\pi)^4\left[i(2\pi)^4 \right]^{-2}\lmd\left[ 2\hat{\zeta}(2-d)\sigma \right]^2 \int d^Dk' \frac{1}{k'^2+m^2}\int d^Dk\frac{1}{k^2+m^2 }\nnb\\
&=&i(2\pi)^{-4}\lmd  4\hat{\zeta}^2\sigma^2\left[ i\pi^2m^2\frac{2-d}{\Gamma(2-d)} \right]^2={\cal J}_a.
\label{mmas_17cc}
\eeqa

We then face another diagram in (d)  due to the trilinear term on RHS of \reflef{mmas2_15});  
%%%%%%%
\beq
{\cal J}_d = \frac{[-i(2\pi)^4]^3}{ [i(2\pi)^4]^{4}}\lmd m^2\left[2\hat{\zeta} (2-d)\sigma \right]^2\left(\frac{3!m}{2\sqrt{3}}\right)^2
\int d^Dk \int d^Dk'\frac{1}{k^2+m^2}\frac{1}{(k'^2+m^2)^2}\frac{1}{(k-k')^2+m^2},
\label{mmas_17c}
\eeq
where the last term under the double integral is  ovelapping not separable into the functions of $k$ alone, and $k'$ alone, respectively, corresponding to the inclined line in (d) in Fig. 1.

We have two alternative ways, carrying out (i) the $k'$ integral first, or (ii) the $k$ integral first.  This also corresponds to including the ovelapping integral as part of $k'$ integral, or $k$ integral, or expressed more explicitly as
%%%%%%%%%
\beqa
{\cal J}_d^{\rm (i)}&=& K_d(2-d)^2\int d^Dk \frac{1}{k^2+m^2}\left( \int d^Dk'\frac{1}{(k'^2+m^2)^2}\:\frac{1}{(k-k')^2+m^2}  \right), \label{mmas_71} \\
{\cal J}_d^{\rm (ii)}&=& K_d(2-d)^2\int d^Dk'\frac{1}{(k'^2+m^2)^2}\left(\int d^Dk \frac{1}{k^2+m^2}\:\frac{1}{(k-k')^2+m^2}  \right),
\label{mmas_72}
\eeqa
where
%%%%%%%%
\beq
K_d=\frac{[-i(2\pi)^4]^3}{[i(2\pi)^4]^4}4\hat{\zeta}^2\sigma^2\left( \frac{m}{2}\sqrt{\frac{\lmd}{3}}3! \right)^2=3i\frac{\lmd m^2}{4\pi^4}\hat{\zeta}^2\sigma^2.
\label{mmas_71ad}
\eeq

We are going to show first that ${\cal J}_d^{\rm (i)}$ vanishes due to the $k'$ integration which does not behave divergently, by analyzing only the relevant portion of \reflef{mmas_71}); 
%%%%%
\beqa
\hat{\cal J}_d^{\rm (i)}&=&(2-d)\int d^Dk'\frac{1}{(k'^2+m^2)^2}\:\frac{1}{(k-k')^2+m^2}=-2 (2-d)\int_0^1x dx\int d^D\tilde{k}'\frac{1}{(\tilde{k}'^2+{\cal M}_{\rm(i)}^2)^3} \nnb\\
&=&-i\pi^2(2-d)\int xdx ({\cal M}_{\rm (i)}^2)^{d-3}\Gamma (3-d)\sim -i\pi^2\frac{2-d}{{\cal M}_{\rm (i)}^2} \rightarrow 0,\quad\mbox{as}\quad d\rightarrow 2,
\label{mmas_71a}
\eeqa
in which the gamma function $\Gamma (3-d)$ is left convergent at $d=2$, hence \reflef{mmas_71a}), with the factor $2-d$, vanishes in the same 
way as in \reflef{mmas2_72}), despite that ${\cal M}_{\rm (i)}^2$ defined by
%%%%%%%%
\beq
{\cal M}_{\rm (i)}^2=m^2+x(1-x)k^2,\quad\mbox{with}\quad \tilde{k}'=k'-k.
\label{mmas_71b}
\eeq
is not a pure constant.

We then move on to ${\cal J}_d^{\rm (ii)}$.  Also using \reflef{mmas2_71}), we find
%%%%%%%%%%
\beqa
{\cal J}_d^{\rm (ii)}&=&K_d(2-d)^2\int_0^1 dx\int d^Dk'\frac{1}{(k'^2+m^2)^2}\int d^D\tilde{k} \frac{1}{(\tilde{k}^2+{\cal M}_{\rm (ii)}^2)^2} \nnb\\
&=&\hspace{-.8em}K_d(2-d)i\pi^2 \int d^Dk'\frac{1}{(k'^2+m^2)^2}({\cal M})_{\rm (ii)}^{d-2}(2-d)\Gamma(2-d) 
\hspace{-.2em}=\hspace{-.2em}K_d(2-d)i\pi^2 \hspace{-.3em}\int d^Dk'\frac{1}{(k'^2+m^2)^2} \nnb\\
&=&K_d(i\pi^2)^2(2-d)\Gamma(2-d)=K_d(i\pi^2)^2= -3i\frac{1}{4\pi^4}\lmd\hat{\zeta}^2m^4 \sigma^2 =3{\cal I}_a,
\label{mmas_73}
\eeqa
3 times as large as the last term of \reflef{mmas2_16}) for ${\cal J}_a$, where 
%%%%%%%%%
\beq
{\cal M}_{\rm (ii)}^2 =m^2 +x(1-x)k'^2, 
\quad\mbox{with}\quad \tilde{k}=k-k',
\label{mmas_74}
\eeq
by the repeated use of CQI.  Notice that the fact that ${\cal M}_{\rm (ii)}^2$ is not purely constant does not affect the conclusion in the limit $d\rightarrow 2$.

Basically the same analysis can be applied to the diagram (e).  From a visual inspection, we readily identify the difference in the equations; we only replace 
 the double-pole terms of the type $(k^2+m^2)^{-2}$ by the single-pole terms
 $(k^2+m^2)^{-1}$, where $k$ might be $k'$.
This removes the difference between \reflef{mmas_71}) and \reflef{mmas_72}), leaving us to consider only \reflef{mmas_73}), with the double-pole term at the very end of the second line replaced by the single-pole term, implying a sign change as we notice between \reflef{mmas2_71}) and \reflef{mmas2_71a}).  This is offset by the difference between $[-i(2\pi)^4]^3/[i(2\pi)^4]^4$ at the top of \reflef{mmas_17c}) and the correponding factor  $[-i(2\pi)^4]^2/[i(2\pi)^4]^3$ supposed to occur in ${\cal J}_e$.  In this way we reach the simple result
%%%%%%%
\beq
{\cal J}_e ={\cal J}_d.
\label{mmas_13}
\eeq

We then take the sum
%%%%%%
\beq
{\cal J}={\cal J}_a +2{\cal J}_b+{\cal J}_c+2{\cal J}_d+2{\cal J}_e =-i\frac{16}{4}\lmd m^4 \hat{\zeta}^2\sigma^2 =-4i\lmd m^4 \hat{\zeta}^2\sigma^2, 
\label{mmas_18}
\eeq
where we have doubled the contribution from (b) and (d), to recover the right-left symmetry, while another doubling in (e) has been applied because the diagrams displayed in Fig. 1 are only half of the two possible choices of the dotted lines.

We are then allowed to  compare our result with the effective mass term of $\sigma$;
%%%%
\beq
{\cal J}=-L_\mu =\half \mu^2 \sigma^2,
\label{mmas_22}
\eeq
thus allowing us to identify \reflef{mmas_18}) with $-i(2\pi)^4 \mu^2\sigma^2$, hence
%%%%%%%%%%
\beq
\mu^2 =\frac{4}{16\pi^4}\hat{\zeta}^2\lmd m^4,\quad\mbox{or}\quad \mu =\frac{1}{2\pi^2}\sqrt{\lmd}\hat{\zeta} m^2.
\label{mmas_23}
\eeq
In this way we have come to conclude that a masless dilaton in the presence of quantum loops does acquire a nonzero mass, thus becoming a pseudo dilaton, almost automatically in accordance with a simple interpretation of the exponential factor in \reflef{mmas2_15}).  It seems even amusing to find that the same Higgs field plays indispensable roles in creating masses both of the pseudo dilaton, part of gravity,  and the rest of the fundamental particles.

In spite of this desired result in deriving the mass $\mu$, we still admit some uncertainty which might arise from the contribution without being  derived from the CQI process.  Consider a typical example of an additional $\tilde{\Phi}$ loop inserted between the two loops in the diagram (a) of Fig. 1, for example, also connected to the neighboring loops through the $\lmd$ term in ${\cal V}$ of \reflef{mmas2_15}).\footnote{Basically the same type of analysis can be applied to another example of a loop attached to the side of the left loop in Fig. 1 (b), for example.}   In the absence of $(d-2)\sigma$ we must appeal to the conventional  procedure in DR;
%%%%%%%%%
\beq
{\cal R}\equiv -i(2\pi)^4\lmd^2 [i(2\pi)^4]^{-2}\int d^Dk\frac{1}{(k^2+m^2)^2}\approx -\frac{\lmd^2}{16\pi^2}\Gamma(-\delta)\exp\left( -\delta /\delta_m\right)
),
\label{mmas_51}
\eeq
where $\delta =d-2,\hspace{.6em} (m^2)^\delta =\exp(\delta \ln m^2) $, and $\lmd \sim {\cal O}(1)$, also  
%%%%%%%%%
\beq
\delta_m \equiv\frac{1}{-\ln m^2} \approx \frac{1}{75}\approx 0.013.
\label{mmas_51a}
\eeq
The foregoing CQI calculations would be left undisturbed if $|\cal R|$ is kept well below the order unity in the reduced Planckian units.

The relation \reflef{mmas_51}) is 
evaluated at $\delta \rightarrow 0+$.  The divergence from $\Gamma (-\delta) \sim -\delta^{-1}$ requires a re-regularization, which is expected to convert a pole to a cutoff $\delta_c>0$, for example.
In the absence of any general way of fixing $\delta_c$, also without related physical observables,\footnote{For the self-masses in QED or QCD, divergences are removed simply to fit the observed values. } we might choose it to be $\delta_m$ defined by \reflef{mmas_51a}), which is  not only unique to the current model with the occurrence of $m^2$, but also happens to be reasonably small.  In this way we have now reached;
%%%%%%%
\beq
|{\cal R}(\delta_c)| = |{\cal R}(\delta_m)|\approx \frac{-\ln m^2}{16\pi^2}e^{-1}\approx \frac{0.48}{2.72}\approx 0.18 \Lsim {\cal O}(1),
\label{mmas_53a}
\eeq
which turns out to be barely a lower end of $ {\cal O}(1)$.

This result achieved by a simple but natural approach  appears to be an encouraging sign for a theoretically favored idea of the CQI-dominance, at least in certain physical situations, though more details are yet to be  scrutinized, probably in the future.  In the next Section, we then enter  the final step of our numerical analysis on the  basis of the expected CQI computation.

\section{Estimating $\mu$}
\setcounter{equation}{0}

In the second equation of \reflef{mmas_23}),  we first re-express $m^2 =(1.26 \times 10^2)^2 ({\rm GeV})^2$ into the Planckian unit system;
%%%%%%%
\beq
m^2 =(1.26 \times 10^2)^2\times  (2.44^{-1} \times 10^{-18})^2.
\label{mmas_23a}
\eeq
However,  $\hat{\zeta}$ has the mass dimension $-1$, so that the product $\hat{\zeta} m^2$ has the mass dimension $+1$, in agreement with $\mu$ on LHS of the second of \reflef{mmas_23}), and is going to be re-expressed in units of GeV instead of $(\rm GeV)^2$;
%%%%%%%
\beqa
\hat{\zeta} m^2&=&\hat{\zeta}(1.26 \times 10^2)^2\times   (2.44^{-1} \times 10^{-18})=\zeta\frac{1.26^2 \times 10^4}{2.44} \times 10^{-18}{\rm GeV} \nnb\\
&=&0.651 \zeta\times 10^{-14}{\rm GeV}=6.51\zeta \;\mu{\rm eV}.
\label{mmas_23b}
\eeqa

In order to know $\lmd$,  we first use \reflef{mmas2_10d}).  We then go through the two well-known steps in SM;
\beq
m_W =-\frac{gv}{2},\quad\mbox{and}\quad \frac{g^2}{m_W^2}=\frac{8}{\sqrt{2}}G_{\rm F},
\label{mmas_31}
\eeq
where $m_W=80.2 {\rm GeV}$ is the mass of the W meson interpreted as a Higgs mechanism for the mass of a gauge field, while $G_{\rm F}=1.17\times 10^{-5}{\rm GeV}^{-2}$ is the Fermi constant for the weak interaction.  We then use the second equation of \reflef{mmas_31}) to derive\footnote{The effect of renormalized fields might have been ignored at this moment, given the crude approximation to be allowed in \reflef{mmas_37}). For details see \cite{sred}, for example.  }
%%%%%%
\beq
g^2 =\frac{8}{\sqrt{2}}m_W^2\: G_{\rm F} =0.426,\quad\mbox{or}\quad g=0.653,
\label{mmas_32}
\eeq
which is then substituted into the first of \reflef{mmas_31}), finding
%%%%%%%
\beq
v=-\frac{2}{g}m_W =246 {\rm GeV},
\label{mmas_33}
\eeq  
finally substituted into \reflef{mmas2_10d})  hence arriving at
%%%%%%%
\beq
\lmd =3\left( \frac{1.26}{2.46} \right)^2= 0.787,\quad\mbox{or}\quad \sqrt{\lmd}=0.887.
\label{mmas_34}
\eeq

We also need an estimate of $\zeta$.  Accordoing to \reflef{nest_2b}), 
we have $\epsilon = -1$ and $1/6 <\xi <1/2$, hence $1/4<\zeta^2 <\infty$ for the radiation-dominated universe, while $1/2<\xi <3/2$ and $3/16 <\zeta^2 <1/4$ for the dust-dominated universe,  as shown in Fig. 1 of \cite{Entr,yf118}. 
Our own fit to the accelerating universe yields $\zeta \sim 1.58$. as read in Fig. 5.8 of \cite{yfkm} and Fig. 9 of \cite{Entr}.  Also noted is $\xi=1/4$ and $\zeta^2 =1/2$, or $\zeta =0.714$ from Super String Theory as indicated in (3.4.58) of \cite{spst}.  In view of these findings, we suggest a tentative but still convenient bound of $\zeta$ somewhere between 0.5 and 2.0;
%%%%%%%
\beq
\zeta\approx (0.5\sim 2.0).
\label{mmas_35}
\eeq

Summarizing them all finally in the second of \reflef{mmas_23}), we obtain
%%%%%%
\beq
\mu \approx (0.15 \sim 0.59)\mu{\rm eV} \sim (150\sim 590){\rm neV}. 
\label{mmas_37}
\eeq
The far RHS suggests nearly two orders of magnitude over the previous estimate $\sim 10^{-9}{\rm eV}$.\footnote{It still appears that the two estimates above are more or less close to each other from a wider point of view, probably because the two approaches share the same concept on the pseudo dilaton in some way or the other. }

\section{Coupling of a pseudo dilaton to two photons} 
\setcounter{equation}{0}

As an application  of the current approach, we now try to re-derive the coupling term
%%%%%%%%%%
\beq
-L_3=A\sigma\frac{1}{4} F_{\mu\nu}F^{\mu\nu},
\label{3p_0}
\eeq
which plays a pivotal role  in the experimental searches for DE \cite{DEexp}, hopefully on a wider perspective in exploiting the scale invariance than in the past attempts \cite{yfkm,Entr}.

To emphasize the unique nature of the pseudo dilaton, we are going to study the photon SE part, or the vacuum polarization, represented by the diagram, the same type as shown on the upper line of Fig. 1, though the solid line, used to be for the neutral Higgs field, is now re-interpreted as a charged matter field, also with the dotted line used for the pseudo dilaton, replaced by the photon line.  For simplicity for the moment, we assume a singly charged and massive Dirac field $\psi$, a representative of  quarks and leptons.

We start off with the simple electromagnetic interaction of $\psi$ first in JF, followed by moving to EF;
%%%%%%
\beq
-{\cal L}_{\rm em}=bie \left(\bar{\psi}b^{i\mu}\gamma_i\psi \right)A_\mu \rightarrow b_*\Omega^{-D}ie \Omega^{D-1}\left( \bar\psi_*b_*^{i\mu}\Omega\gamma_i\psi_* \right)\Omega^{d-2}A_{*\mu}\equiv -b_*\Omega^{d-2}L_{\rm *em},
\label{3p_1}
\eeq
where $e$ is the elementary charge chosen to be a pure constant, with $\alpha =e^2/(4\pi)\approx 1/137$;
%%%%%%%%%%
\beq
-L_{\rm *em} =ie \bar{\psi}_* b^{i\mu}_*\gamma_i\psi_* A_{*\mu}, 
\label{3p_2}
\eeq
with $b_\mu^i$ the dimensionally extended tetrad with $b=\sqrt{-g}$, also the electromagnetic field transforming as $A_\mu =\Omega^{d-2}A_{*\mu}$. The occurrence of $\Omega^{d-2}$ on the far RHS of \reflef{3p_1}) indicates scale invairance in $4$ dimensions, hence the same nature as generating $\sigma$  as in the previous Sections.

In EF, we approximate spacetime by locally Minkowskian to apply the ordinaly Feynman rules based on \reflef{3p_2}) to the same type of the diagram as on the upper line of Fig. 1 in terms of DR, obtaining the gauge-invariant form\footnote{See (6.174)-(6.181) of \cite{yfkm} for details simply for the scalar loop field.  Extending to the Dirac field is tedious but straightforward.}
%%%%%%%%%
\beq
\Pi_{\*\mu\nu}(k)=-\frac{\alpha}{3\pi}\left( k_\mu k_\nu -k^2\eta_{\mu\nu} \right)\Gamma(2-d),
\label{3p_3}
\eeq 
where we have reversed the overall sign due to the antisymmetric nature of $\psi$ and $\bar{\psi}$.

Further re-installing $\Omega^{d-2}$ on the far RHS of \reflef{3p_1}), substituted from the third of \reflef{nest_2}), we reach the whole result
%%%%%
\beq
\sqrt{-g_*}\exp\left( 2\hat{\zeta}(d-2) \sigma \right)\frac{-\alpha}{3\pi}\left( k_\mu k_\nu -k^2\eta_{\mu\nu} \right)\Gamma(2-d).
\label{3p_4}
\eeq 
We then pick up the term liner in $\sigma$, also using the CQI in the form of \reflef{mmas2_61b}), comparing the result with \reflef{3p_0}) by using the relation
%%%%%%%
\beq
\frac{1}{4}F_{\mu\nu}F^{\mu\nu}=-\epsilon_f^\mu \left( k_\mu k_\nu-k^2\eta_{\mu\nu} \right) \epsilon_i^\nu,
\label{3p_6}
\eeq
with $\epsilon_f^\mu$ and $\epsilon_i^\nu$ for the polarization vectors of the final and initial photons, respectively, then identifying the constant $A$ as
%%%%%%%%
\beq
A=-\frac{\alpha}{3\pi}\hat{\zeta}.
\label{3p_5}
\eeq
In this way we come to determine $A$ basically of the size of the inverse of the Planck mass, as expected to be.

For each of the quarks and the  leptons, the result \reflef{3p_5}) adds up with the corresponding multiplicative factors for the electric charges squared.  Notice also that a cancellation always takes place between the terms of the same $m$, the mass of the loop fields, to meet the gauge-invariance of the form \reflef{3p_3}), thus yielding $A$ independnent of $m$.  This has an advantege that we derive $A$ basically $\sim \alpha\hat{\zeta}$, but, on the other hand, deprives us of a familiar procedure  to suppress the contribution from  much heavier and uncertain loop fields.

From this point of view, we may also consider the charged scalar fields,  sharing the same charge-structure as indicated by supersymmetry, for which we develop obviously the pararell computations with $\psi$ in \reflef{3p_2}) replaced by $\Phi$ together with the additional 4-point term $\sim e^2\bar\Phi\Phi g^{\mu\nu} A_\mu A_\nu$,\footnote{This term contributes a term proportional to $m^2$ due to a simple 1-loop of $<0|\Phi\bar{\Phi}|0> \propto m^2$ in DR , thus cancelling the term of $\sim m^2$ in the main loop term as in the upper line of Fig. 1.} without the sign change due to the antisymmetry of the fermionic field, ending up with a multiplicative factor $-1/4$ to \reflef{3p_3}), 
%%%%%%%%%%
\beq
A_{\rm sc}=\frac{\alpha}{12\pi}\hat{\zeta}.
\label{3p_7}
\eeq   
in place of \reflef{3p_5}).

The reduction factor $1/4$ can be interpreted by $(1/2)^2$ with 1 and 2  
 for the spin degrees of freedom for the scalar and the Dirac fields, respectively, while the squaring takes care of the occurrence of two lines in each of the main loop diagrams.  In this sense, every 4 scalar fields offset the effect of 1 Dirac field, no matter how heavy they might be. Obviously, it appears too early to make an unambiguous prediction before we develop a more general survey on what the fundamental particles are to be included in the loop, also considering wider class of spin-statistics combinations,  again left to future studies, at this time.  An uncertainty of this kind still unavoidable at present might result in an adjustable paramter multiplied  to $\Gamma^{1/2}$ with $\Gamma$ for the decay width of $\sigma$ into two photons in the formulation in \cite{DEexp}.\footnote{The previous result Eq. (91) in \cite{Entr}, for example, can even be re-interpreted as our \reflef{3p_5}) multiplied by an ``adjustable parameter'' $B/A=\zeta^{-1}(2/3){\cal Z}$.}

\section{Summary}
\setcounter{equation}{0}

We started out by assuming the scale-invariance of the Higgs potential in JF in STT in 4 dimensions, reaching, somewhat unexpectedly, a time-independent particle mass in EF in conformity with today's view on measuring cosmological size in units of the inverse of the mass of the microscopic particles.   We then extended the spacetime dimensionality $D$ off the physical value 4, allowing us to analyze the behavior of the pseudo NG boson, pseudo dilaton.  We derived its mass by studying the $\sigma$ SE part.  By considering the 2-loop amplitudes, we obtained the nonzero and finite mass $\mu$ of the pseudo dilaton, by maximally exploiting the CQI relation, also utilizing SM, somewhere around $\mu{\rm eV}$, which turns out approximately 2 orders of magnitude heavier than our previous tentaitve  estimate.
The difference is understood naturally because we now deal with a theoretical model quite different from our previous simple-minded one.  It still seems helpful if we find any aspect more tractable on the non-CQI terms possbily in the future.
 As an extended idea, we tried also to re-derive the coupling strength of $\sigma$ into $2\gamma$, still short of the fully unique determination of the multiplier at present.

\mbox{}\\[.5em]

\noindent
{\Large\bf Acknowledgments}
\mbox{}\\[.2em]

The author expresses his sincere thanks to K. Homma, H. Itoyama,  C.S. Lim, K. Maeda, T. Tada and T. Yoneya for many useful discussions.

%%%%%%%%%%%%%%%%% ref  %%%%%%%%%%%%%%%%%%%%%
\small{

}

\end{document}